\documentclass{article}
\usepackage{pstricks,graphics}

\begin{document}

\title{Dynamic Memory Management on GPUs with SYCL}
\author{Russell K. Standish\\High Performance Coders}

\maketitle
\begin{abstract}
  Dynamic memory allocation is not traditionally available in kernels
  running on GPUs. This work aims to build on Ouroboros, an efficient
  dynamic memory management library for CUDA applications, by porting
  the code to SYCL, a cross-platform accelerator API. Since SYCL can
  be compiled to a CUDA backend, it is possible to compare the
  performance of the SYCL implementation with that of the original
  CUDA implementation, as well as test it on non-CUDA platforms such
  as Intel's Xe graphics.
\end{abstract}

\section{Introduction}

Dynamic memory allocation is not traditionally part of the runtime
environment of code running on accelerators, such as {\em graphical
  processing units} (GPUs), {\em signal processors} (DSPs) and {\em field
  programmable gate arrays} (FPGAs). Instead memory is allocated by
the host code to fixed sizes prior to launching the kernel code that
runs on the accelerator. This works well for many applications with
fixed partitioning of space and time, such as finite element methods
of partial differential equations, or algorithms based on matrix
multiplication.

However, some applications, such as graph algorithms, or agent based
models, require memory to be dynamically partitioned between the
objects of the computation. Stopping the kernel, resizing memory
allocations and relaunching is simply unfeasible in those
circumstances. Instead, what is needed is to preallocate a chunk of
memory on the host to act as a {\em heap}, and to run a heap
allocation algorithm on the accelerator device.

There are two main C++ APIs for programming the host and kernel
code. CUDA\cite{luebke2008cuda} is an extension to standard C++, and
kernel code and host code are distinct. CUDA is developed by NVIDIA,
and supports only GPUs manufactured by that company. SYCL\cite{reinders2023data},
on the other hand, does not distinguish between host and kernel code,
it is all standard C++. However, not all standard library features are
available in a kernel context, which will be flagged by the
compiler. Drivers are available for all major GPU types from Intel,
AMD and NVIDIA, so SYCL is considered cross-platform.

At this point, it is worth mentioning a third option, also
cross-platform, called OpenCL.  It is an older C API, and
consequentially less feature rich, and more difficult to use than CUDA
or SYCL.

CUDA gained dynamic memory allocation in 2009\cite{CUDAguide},
but is often considered slow and unreliable. Since then, a number of
dynamic memory allocators have been proposed, mostly for CUDA,
although one non-open source OpenCL implementation
exists\cite{spliet2014kma}. Details of these can be found in a survey by Winter
and Mlakar\cite{winter2021dynamic}. The basic strategy is to divide the
preallocated block into chunks of different sizes, each of which go
into a lock-free queue dedicated for the particular chunk size.

In the above survey, benchmarks indicate Ouroboros\cite{winter2020ouroboros}, by
the same authors, as being the most performant. Therefore this work
took the CUDA Ouroboros code and translated it into SYCL, called
Ouroboros-SYCL\footnote{https://github.com/highperformancecoder/Ouroboros-SYCL}.
Since SYCL has implementations that target CUDA, we can directly
compare the resulting performance of Ouroboros-SYCL with the original
Ouroboros code on the same NVIDIA hardware, as well as provide results
on non-NVIDIA hardware (Intel Xe graphics).

\section{Porting the code to SYCL}

\noindent As a way of getting started, I attempted to use the automatic CUDA to
SYCL translator, called {\em SYCLomatic}\cite{liang2024using}. As might be
expected for a project with the complexity of Ouroboros, this tool
failed to generate compilable SYCL code, but managed to convert
perhaps half of the code into SYCL equivalents, leaving constructs it
failed to convert in the original CUDA form, and annotating some other
parts it felt need further attention from the programmer to check the
conversion was valid. In the end, whilst the code conversion was a
good start, pretty much all of the automatically converted code was
modified manually. The first reason for this is that CUDA always uses a
3D layout of processing elements, whereas SYCL allows for the
possibility of 1, 2 or 3D layouts of processing elements, indicated by
a template parameter. SYCLomatic always renders the translated code as
3D, to match the original source code, however, Ouroboros, being a
library needs to handle arbitrary layouts. Thus code to extract, for
example, a processing element's rank (known as {\em id} in SYCL) uses
the \verb+get_global_linear_id()+ call, which returns a single number
regardless of the dimensionality of the processing element layout.

In the process of getting the code to compile, SYCLomatic failed to
convert atomic memory operations. SYCL has an atomic reference type,
that wraps a variable, and allows a range of atomic operations,
including the usual binary operation suspects
($+,-,\wedge,\vee,\oplus$,min,max), compare\_and\_swap. CUDA, by contrast
has a set of atomic library functions, implementing the same
operations. In the end, the simplest approach was to provide
implementations of the CUDA library functions, implemented using SYCL
atomic reference types. 

The next issue was how to represent CUDA's global \verb+threadIdx+ and
\verb+blockIdx+ variables. These refer to the coordinates (in CUDA's
3D space) of the thread (within its block) and the coordinates of the
block. In SYCL terminology, a CUDA block is a {\em group}. An
\verb+nd_item+ object contains all the information about the current
thread's rank within its group, and its group's rank within the
global range. However, the \verb+nd_item+ object is not available
as a global function, but must be passed as a parameter into the stack
frame where it is needed. A similar issue relates to I/O --- CUDA has a
\verb+printf+ statement available that is callable in any kernel function,
whereas the SYCL equivalent is a \verb+stream+ object, usable like
\verb+std::cout+ that must be created in the command group scope, and
passed passed as a parameter to inner function scopes. In the
Ouroboros-SYCL case, each class member function called in kernel code
must take a template parameterised parameter (\verb+Desc+) that has an
item (an \verb+nd_item+ of arbitrary rank) and an out object that supports
\verb+operator<<+ serialisation. Being a template parameter, users of
the library can choose to define the rank of item, and use a
\verb+sycl::stream+ or a dummy out object as appropriate.

It should be noted that Intel's oneAPI SYCL compiler (called DPC++)
also provides an experimental free function \verb+get_nd_item()+, and
an experiment free function \verb+printf()+ that can be used for this
purpose. These functions are proposed for a future SYCL standard.

Another point about the \verb+sycl::stream+ object is that it buffers
the string data written to it, and the message is only written to the
console when the stream object goes out of scope. Unfortunately, if
the problem being diagnosed is a deadlock, or a crash, the stream
object never goes out of scope, so any helpful debug messages written by way
of this object will not be seen --- a frustrating exercise indeed.

CUDA compute capability 7 introduced a \verb+nanosleep+ function for pausing
threads for a specific period of time. Ouroboros uses this function to
throttle threads demanding to allocate memory so that other threads
freeing memory can catch up. This function would indeed be a useful
optimisation technique, but is unavailable in the SYCL programming
environment. Instead, all we can do is perform an
\verb+atomic_fence()+, which ensures that other threads catch up to
the fence.

The final difficulty in converting the Ouroboros CUDA code had to do
with the use of warp vote functions, which allowed multiple
allocations to occur within one warp. In SYCL terminology, a warp is
known as a {\em subgroup}, and corresponding group reduction
algorithms exist, applied to subgroups, that are the equivalent of the
warp functions. Unfortunately there is an issue. The CUDA equivalents
take a mask, so that threads not participating in the group operation
can be masked out by passing the results of \verb+__activemask()+.

In SYCL, there is no real way of obtaining the active mask, and
according to the standard, group operations block until all threads
call the group operation function.

It would be more useful if group operations require all subgroups
within a group to participate, and only those active threads should be
required to call the group operation. So it should be possible to
obtain the active mask by means of the following code:

\begin{verbatim}
auto sg=i.get_sub_group();
auto activeMask=sycl::reduce_over_group(
        sg,
        1ULL<<sg.get_local_linear_id(),
        sycl::bit_or<>()
    );
\end{verbatim}

Interestingly, when run on an Intel GPU, or on the CPU, this code runs
as expected, and generates the active mask. But when run on an NVIDIA
GPU, this code deadlocks, both with Intel's oneAPI, and with the
AdaptiveCpp compiler, unless all threads in the subgroup are active.

\section{Methods}

\noindent Ouroboros comes with a driver program for each of the six alternative
heap algorithms, {\em chunk}, {\em page}, {\em virtual array chunk},
{\em virtual array page}, {\em virtual list chunk} and
{\em virtual list page}. Arguments passed to the driver program
specify the data size to be allocated, and number of allocations to be
allocated in parallel. Finally, the program iterates ten times through
allocating memory, writing some data, checking that the data is correct when
read back and then freeing the memory.  The average time for
performing the allocations and frees is calculated.

In this work, one additional change was made to the original code,
aside from a trivial change to reduce the total amount of heap space
available in order to fit on device available to the author. SYCL
implementations typically compile the kernel code into an intermediate
representation, such as SPIR-V\cite{kessenich2018spir}, and transpiling this into
the native machine code of the accelerator occurs in a {\em just in
  time} (JIT) fashion when the kernel is first launched. As a result,
there can be a big disparity between the time recorded for the first
iteration, and the times recorded for subsequent iterations. So the
code was modified to report the average over all iterations, and the
average over all but the first iteration (ie {\em subsequent
  iterations}). This allow a more apples-to-apple comparison between
the CUDA implementation and the SYCL implementation.

The actual code can be found in the Ouroboros-SYCL GitHub
repository\footnote{https://github.com/highperformancecoder/Ouroboros-SYCL}. The
SYCL code is available in the {\em master} branch, and the original
optimised CUDA code in the {\em cuda-ouroboros} branch.

\begin{description}
\item[Hardware:]\mbox{}\\
  \begin{enumerate}
  \item Dell Precision 7540 laptop with i9-9880H CPU @ 2.3GHz and NVIDIA
    Quadro T2000 GPU.
  \item Asus NUC 13, i5-1340P CPU with integrated Iris Xe graphics GPU.
  \end{enumerate}

\item[Software:]\mbox{}\\
  \begin{enumerate}
  \item Intel oneAPI 2025.1 (icpx compiler)
  \item Codeplay's oneAPI for NVIDIA GPUs plugin
  \item CUDA 12.8
  \item Adaptive C++, compiled from source
    code\footnote{https://github.com/AdaptiveCpp/AdaptiveCpp}, commit
    f336ab84.  Adaptive C++ was previously known as HipSYCL.
  \end{enumerate}
\end{description}

After running cmake, it is necessary to insert the compiler manually
using ccmake: for oneAPI the compiler is {\tt icpx}, and you need to
add the option {\tt -fsycl}, as well as for NVIDIA use, the option
{\tt -fsycl-targets=nvptx64-nvidia-cuda}. For Adaptive C++, the
compiler is acpp, and doesn't require any special command line flags.

The use of these compilers allows the comparison of the translated code
with original code, running on the same hardware. Adaptive C++ targets
CUDA's PTX machine, so is closer to what the CUDA compiler nvcc
produces. However, it does the final of intermediate code to ptx in a {\em
  just-in-time} fashion, so for a proper comparison, we should compare
only measured alloc/free times after the first iteration. Similarly,
Codeplay's plugin performs JIT compilation of intermediate code to
ptx.

The optimised Ouroboros code has a few instances of embedded PTX code,
also making use of \verb+nanosleep()+, and the ability to mask warp
voting functions by the active mask. To make the comparison fair with
the SYCL versions, I created a {\em deoptimised version}, with the
embedded code replaced by high level code equivalents, \verb+nanosleep+
replaced by an \verb+atomic_fence+, and the code using warp functions
replaced by the simplified code used in the SYCL versions. This code
can be found in the {\em deoptimised} branch of the Ouroboros-SYCL repository.

\section{Results}

\noindent In interpreting the algorithm results, it should be noted that the
heap is divided into {\em chunks} of different sizes, and the
allocation requests are served as {\em pages} from within each chunk.

Note that the Adaptive C++ compiled code would struggle as the number
of threads increased, with loops timing out or becoming deadlocked.

The raw results files are available in the supplementary
materials\cite{Standish25a}, and a Ravel\footnote{Ravel is a
  revolutionary product for interactively analysing multidimensional
  data, available from https://ravelation.net} file with the data
loaded to assist in the data analysis.

\subsection{Page allocator}

\begin{figure}
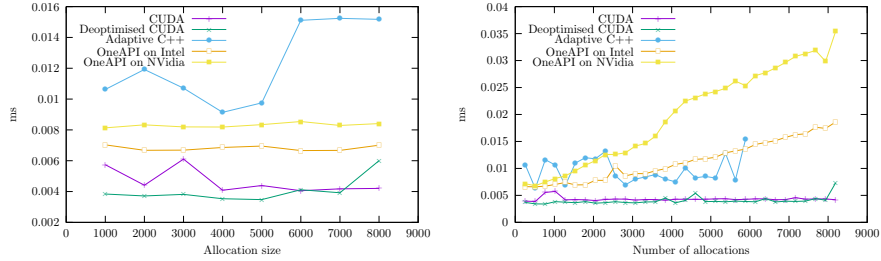

  \resizebox{0.48\textwidth}{!}{\input{suballoc-page-numalloc=1024}}
  \resizebox{0.48\textwidth}{!}{\input{suballoc-page-size=1000}}
  \caption{Average subsequent allocation time of the page allocator
    as a function of allocation size for 1024 allocations, and as a
    function of number of simultaneous allocations for an allocation
    size of 1000 bytes}
  \label{page-allocator}
\end{figure}

The simplest allocator is the page-based allocator, where pages of
fixed size are allocated from a queue. Total heap memory is divided
amongst the queues, each queue managing a different page size. Being
the simplest allocator, it is also the fastest, but suffers more from
fragmentation than the other more sophisticated schemes. Figure
\ref{page-allocator} shows the average subsequent timings of
allocations as a function of allocation size when 1024 threads are
attempting to simultaneously allocate, and the timings as a function
of number of threads simultaneously attempting to allocate 1000 bytes.

The performance of the SYCL code ends up being about half that of the
CUDA code. Interestingly, the attempt to deoptimise the CUDA code to
make it more comparable to the SYCL version only seem to make it more
performant, if anything.

\subsection{Chunk allocator}

\begin{figure}
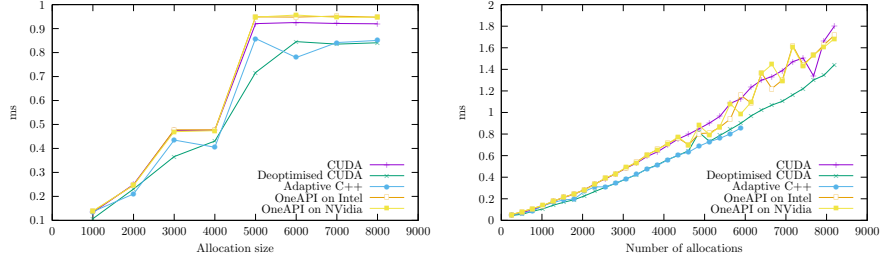

  \resizebox{0.48\textwidth}{!}{\input{suballoc-chunk-numalloc=1024}}
  \resizebox{0.48\textwidth}{!}{\input{suballoc-chunk-size=1000}}
  \caption{Average subsequent allocation time of the chunk allocator
    as a function of allocation size for 1024 allocations, and as a
    function of number of simultaneous allocations for an allocation
    size of 1000 bytes}
  \label{chunk-allocator}
\end{figure}

The chunk allocator maintains queues of chunks that have free pages,
first obtaining a chunk index, then scanning the chunk for free
pages. It is a more complex algorithm, but queue sizes are smaller.

Figure \ref{chunk-allocator} (left) shows the average time to allocate
memory for different allocation sizes. The allocator is implemented as
a linked list of chunk queues, each queue managing chunks sized
according to powers of two. You can see the effect of having to walk
through this link list as the chunk size increases. On the right, you
can see the effect of thread contention as more threads attempt to
allocate chunks simultaneously. We can conclude from these figures
that not only does the SYCL version work (data is written to the
allocated chunks and checked), but that the implementation performance
is broadly in line with the original Ouroboros implementation when run
on the same hardware.

\subsection{Virtualised array and list allocators}

\begin{figure}
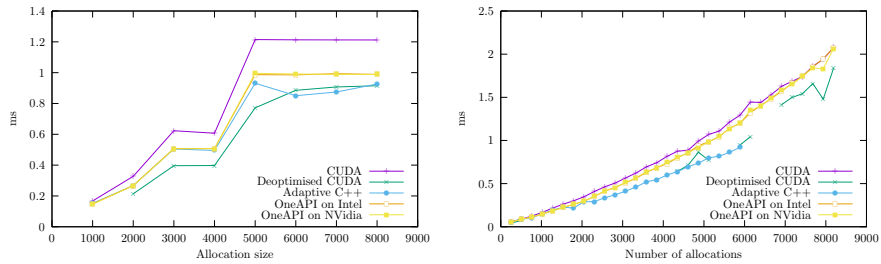

  \resizebox{0.48\textwidth}{!}{\input{suballoc-vl-chunk-numalloc=1024}}
  \resizebox{0.48\textwidth}{!}{\input{suballoc-vl-chunk-size=1000}}
  \caption{Average subsequent allocation time of the virtualised list
    chunk allocator as a function of allocation size for 1024
    allocations, and as a function of number of simultaneous
    allocations for an allocation size of 1000 bytes}
  \label{vl-chunk-allocator}
\end{figure}

Ouroboros also introduces virtual queues, which reduce queue sizes
even further. Figure \ref{va-page-allocator}-\ref{vl-chunk-allocator}
shows the equivalent results for the virtualised versions of the page
and chunk allocators.

\begin{figure}
  \resizebox{0.48\textwidth}{!}{\input{suballoc-va-page-numalloc=1024}}
  \resizebox{0.48\textwidth}{!}{\input{suballoc-va-page-size=1000}}
  \caption{Average subsequent allocation time of the virtualised array
    page allocator as a function of allocation size for 1024
    allocations, and as a function of number of simultaneous
    allocations for an allocation size of 1000 bytes}
  \label{va-page-allocator}
\end{figure}

\begin{figure}
  \resizebox{0.48\textwidth}{!}{\input{suballoc-vl-page-numalloc=1024}}
  \resizebox{0.48\textwidth}{!}{\input{suballoc-vl-page-size=1000}}
  \caption{Average subsequent allocation time of the virtualised list
    page allocator as a function of allocation size for 1024
    allocations, and as a function of number of simultaneous
    allocations for an allocation size of 1000 bytes.}
  \label{vl-page-allocator}
\end{figure}

\begin{figure}
  \resizebox{0.48\textwidth}{!}{\input{suballoc-va-chunk-numalloc=1024}}
  \resizebox{0.48\textwidth}{!}{\input{suballoc-va-chunk-size=1000}}
  \caption{Average subsequent allocation time of the virtualised array
    chunk allocator as a function of allocation size for 1024
    allocations, and as a function of number of simultaneous
    allocations for an allocation size of 1000 bytes}
  \label{va-chunk-allocator}
\end{figure}

\section{Conclusion}

\noindent The results indicate that the conversion of Ouroboros's CUDA-based
code into SYCL was successful, and within a factor of 2 performance of
the original code for the faster page-based algorithms, and within
statistical noise of the performance of the chunk-based algorithms
using Intel's oneAPI toolset. Adaptive C++ unfortunately suffered from
timeouts and deadlocks, which may limit the use of this code with this
compiler. As it hasn't yet fully implemented the SYCL 2020 standard,
perhaps this is a matter of time.

The exercise also highlighted some deficiencies of SYCL with respect
to CUDA --- in particular the need for global access to a thread's
\verb+nd_item+, a global printf function for debugging purposes (both
of these are proposed as experimental additions to SYCL in the oneAPI
toolset) and the need for group reduction algorithms to be masked by the
active threads only.

\bibliographystyle{plain}
\bibliography{rus}

\end{document}